\documentstyle[12pt]{article}
\title{Dynamic properties
       of spin-$\frac{1}{2}$ $XY$ chains}
\author  {Oleg Derzhko and Taras Krokhmalskii\\
  \small {\em Institute for Condensed Matter Physics,}\\
  \small {\em 1 Svientsitskii St., L'viv-11, 290011, Ukraine}}
\date{\today}

\begin{document}

\maketitle
\begin{abstract}
{We have considered a numerical scheme 
for the calculation
of the equilibrium properties of
spin-$\frac{1}{2}$ $XY$ chains. 
Within its frames it is necessary to solve
in the last resort only 
the $2N\times 2N$ eigenvalue and eigenvector problem
but not the $2^N\times 2^N$ one 
as for an arbitrary system
consisting of $N$ spins $\frac{1}{2}$.
To illustrate the approach 
we have presented some new results.
Namely,
the $xx$ dynamic structure factor for the Ising model in transverse field,
the density of states for the isotropic chain with
random intersite couplings and transverse fields 
that linearly depend on the surrounding couplings,
and the $zz$ dynamic structure factor
for the Ising model in random transverse field.
The results obtained
are hopped to be useful for an interpretation of observable data 
for one-dimensional spin-$\frac{1}{2}$ $XY$ substances.}
\end{abstract}

\vspace{1cm}


\vspace{1cm}


\vspace{1mm}

\noindent
{\bf Postal addresses:}\\
\em
Dr. Oleg Derzhko\\
Dr. Taras Krokhmalskii\\
Institute for Condensed Matter Physics\\
1 Svientsitskii St., L'viv-11, 290011, Ukraine\\
Tel: (0322) 42 74 39

\noindent
\hspace{7mm}
(0322) 76 09 08\\
Fax: (0322) 76 19 78\\
E-mail: derzhko@icmp.lviv.ua

\noindent
\hspace{14mm}
krokhm@icmp.lviv.ua

\clearpage

\renewcommand\baselinestretch {1,3}
\large\normalsize

Statistical mechanics of
the one-dimensional spin-$\frac{1}{2}$ $XY$ models
has a history going back more than three decades to the pioneering
paper by Lieb, Schultz and Mattis [1]
in which it was noted
that such systems are as a matter of fact the non-interacting
spinless fermions and therefore,
a lot of statistical mechanics
calculations can be performed exactly.
Although many papers concerning thermodynamics, spin
correlations and their dynamics have appeared
since that time,
several problems
still call for efforts.
One of them regards the study of spin dynamics:
in contrast to the time-dependent
correlation function between $z$-components of two spins [2], the
$xx$ time-dependent correlation function has been derived exactly only in
some limiting cases
($T=0$, the critical value of transverse field, $T=\infty$)
[3-9].
Some recent attempts to calculate this correlation function are
presented in Refs. [10-14].
New difficulties arose when
the one-dimensional spin-$\frac{1}{2}$ $XY$ models
were started
to be discussed in random versions. The analytical results
obtained here are not so impressive as for the perfect case and they are
restricted to special types of disorder [15-20] or
renormalization-group analysis [21].
A discovery of the quasi-one-dimensional spin-$\frac{1}{2}$
systems
(Cs(H$_{1-{\mbox{x}}}$D$_{\mbox{x}}$)$_2$PO$_4$,
PbH$_{1-{\mbox{x}}}$D$_{\mbox{x}}$PO$_4$,
PrCl$_3$,
CsCuCl$_3$,
CsCu$_{1-{\mbox{x}}}$Mn$_{\mbox{x}}$Cl$_3$,
$J$-aggregates etc.)
gave arise to additional interest in such calculations
since a lot of data obtained in the dynamic experiments
for such materials are available.
Some approximate studies
of dynamic properties
were inspired by corresponding measurements [22-27],
however, such estimations contain
uncontrolled mistakes.
Additional interest in such calculations was caused by the very recent
studies of quantum phase transitions in disordered systems [21,28-30].

Since there are notorious difficulties in the analytical study of some
properties
of the spin-$\frac{1}{2}$ $XY$ chains
it is naturally to try to obtain the desired
results numerically. Few earlier attempts [31,32]
faced the $2^N\times 2^N$ eigenvalue and eigenvector problem
that restricted the computations to $N\sim 10$.
However,
a peculiarity of spin-$\frac{1}{2}$ $XY$ chains
provides
an evident possibility to compute all equilibrium quantities
facing
in the last resort only
the $2N\times 2N$ eigenvalue and eigenvector problem [33-35].
This fact
allows to consider rather long chains
($N\sim 100-10000$)
and therefore to study reliably
the dynamics of spin correlations 
or the influence of disorder on observable
properties.
In this paper we shall briefly explain such a numerical approach
for calculation of the equilibrium properties of
the spin-$\frac{1}{2}$ $XY$ chains. 
Besides we shall demonstrate how does the approach work
computing for this purpose 
the $xx$ dynamic structure factor
of the transverse Ising model
and some thermodynamic and
dynamic properties of several random $XY$ models.

We shall consider $N$ spins one-half arranged in a row with the
following Hamiltonian
\begin{eqnarray}
H=\sum_{j=1}^{N}\Omega_js_j^z+
\sum_{j=1}^{N-1}
\sum_{\alpha ,\beta =x,y}
J^{\alpha \beta}_js^{\alpha}_js^{\beta}_{j+1}
\end{eqnarray}
where $\Omega_j$ is the transverse field at site $j$
and $J_j^{\alpha \beta}$
is the interaction between $\alpha$ and $\beta$ spin components at
the sites $j$ and $j+1$.
Introducing instead of the spin raising and lowering operators
via the Jordan-Wigner transformation
the Fermi operators
$c^+_j$, $c_j$
one finds that the Hamiltonian (1) is a bilinear fermion form that
can be put into diagonal form by linear transformation
$\eta_k^+=\sum_{j=1}^N\left( h_{kj}^*c_j+g_{kj}^*c_j^+\right) $.
Similarly to [1] it can be shown that if
\begin{eqnarray}
({\bf g}_k,{\bf h}_k)
\left(
\begin{array}{cc}
{\bf A}   &{\bf B}\\
-{\bf B}^*&-{\bf A}^*
\end{array}
\right)
=\Lambda_k({\bf g}_k,{\bf h}_k)
\end{eqnarray}
with
${\bf g}_{k} \equiv (g_{k1},...,g_{kN})$,
${\bf h}_{k} \equiv (h_{k1},...,h_{kN})$,
$A_{ij} \equiv \Omega_i \delta_{ij}+ J^{+-}_i \delta_{j,i+1}+
J^{-+}_{i-1} \delta_{j,i-1}$,
$B_{ij} \equiv J^{++}_i \delta_{j,i+1} -J^{++}_{i-1}\delta_{j,i-1}$
and
$J^{+-}_j \equiv
\frac{1}{4}[J^{xx}_j+J^{yy}_j+{\mbox{i}}(J^{xy}_j-J^{yx}_j)]
={J^{-+}_j}^*$,
$J^{++}_j \equiv
\frac{1}{4}[J^{xx}_j-J^{yy}_j-{\mbox{i}}(J^{xy}_j+J^{yx}_j)]$
the Hamiltonian transforms into
$H=\sum_{k=1}^N\Lambda_k\left( \eta^+_k\eta_k-\frac{1}{2}\right)$.
Evidently, the knowledge of
$\Lambda_k$s or their distribution
$\rho (E)\equiv \frac{1}{N}\sum_{k=1}^N \delta (E-\Lambda_k)$
yields
thermodynamics of the spin system (1).
For the calculation of spin correlation functions
it is convenient to introduce
the auxiliary operators
$\varphi^{+}_j \equiv c_j^{+}+ c_j
= \sum_{p=1}^N
\left( \Phi_{pj}\eta^+_p
+\Phi_{pj}^*\eta_p\right)$
and
$\varphi^{-}_j \equiv c_j^{+}- c_j
= \sum_{p=1}^N
\left( \Psi_{pj}\eta^+_p
-\Psi_{pj}^*\eta_p\right)$,
where
$\Phi_{pj} \equiv g_{pj}+h_{pj}$,
$\Psi_{pj} \equiv g_{pj}-h_{pj}$.
Since
$
s_j^x= \frac{1}{2}
\varphi_1^+ \varphi^-_1
\ldots
\varphi_{j-1}^+ \varphi^-_{j-1}
\varphi^+_j$,
$
s_j^y= \frac{1}{2{\mbox{i}}}
\varphi_1^+ \varphi^-_1
\ldots
\varphi_{j-1}^+ \varphi^-_{j-1}
\varphi^-_j$
and
$
s_j^z=-\frac{1}{2} \varphi_j^+ \varphi^-_j$,
the calculation of thermodynamic
average
of a product of spin operators
reduces to exploiting
the Wick-Bloch-de Dominicis theorem
with the result that 
is in fact the
Pfaffian of corresponding antisymmetric matrix
constructed from the elementary contractions
\begin{eqnarray}
\langle\varphi^+_j(t) \varphi^+_m\rangle=
\sum_{p=1}^N \left[
\frac{
\Phi_{pj}\Phi_{pm}^*}
{{\cal {F}}(\Lambda_p)}
+
\frac{
\Phi_{pj}^*\Phi_{pm}}
{{\cal {F}}(-\Lambda_p)}
\right],
\nonumber\\
\langle\varphi^+_j(t) \varphi^-_m\rangle=
\sum_{p=1}^N \left[
-
\frac{
\Phi_{pj}\Psi_{pm}^*}
{{\cal {F}}(\Lambda_p)}
+
\frac{
\Phi_{pj}^*\Psi_{pm}}
{{\cal {F}}(-\Lambda_p)}
\right],
\nonumber\\
\langle\varphi^-_j(t) \varphi^+_m\rangle=
\sum_{p=1}^N \left[
\frac{
\Psi_{pj}\Phi_{pm}^*}
{{\cal {F}}(\Lambda_p)}
-
\frac{
\Psi_{pj}^*\Phi_{pm}}
{{\cal {F}}(-\Lambda_p)}
\right],
\nonumber\\
\langle\varphi^-_j(t) \varphi^-_m\rangle=
-\sum_{p=1}^N \left[
\frac{
\Psi_{pj}\Psi_{pm}^*}
{{\cal {F}}(\Lambda_p)}
+
\frac{
\Psi_{pj}^*\Psi_{pm}}
{{\cal {F}}(-\Lambda_p)}
\right]
\nonumber
\end{eqnarray}
with
${\cal {F}}(x) \equiv
\left( 1+{\mbox {e}}^{\beta x}\right)
{\mbox {e}}^{-{\mbox{i}}xt}$.
Thus, the solution of the $2N\times 2N$ eigenvalue and eigenvector
problem (2)  completely determines
thermodynamics, spin correlations and their
dynamics for the model (1).

In what follows we have collected
several new results 
obtained within the frames of the described approach.
We shall start from the $xx$ dynamic structure factor
$S_{xx} (\kappa ,\omega )
\equiv
\sum_{n=1}^N{\mbox {e}}^{{\mbox{i}}\kappa n}
$
$
\int_{-\infty}^{\infty}{\mbox{d}}t
{\mbox {e}}^{-\varepsilon \mid t\mid}
{\mbox {e}}^{{\mbox{i}}\omega t}
\langle s_j^{x}(t)s_{j+n}^{x}\rangle$
for the uniform transverse Ising chain
($\Omega_j=\Omega=0.2$,
$J_j^{xx}=J=-1$,
$J_j^{xy}=J_j^{yx}=J_j^{yy}=0$).
At first
we solved the eigenvalue and eigenvector problem (2)
for $N=280$
obtaining in result
$\Lambda_k$, $\Phi_{kj}$, $\Psi_{kj}$.
Then for several values of temperature 
we calculated the required elementary contractions
and computed the relevant Pfaffian
\begin{eqnarray}
\langle s_{32}^x(t)s_{32+n}^x\rangle=
\frac{1}{4}
\langle\varphi_1^+(t)
\varphi_1^-(t)
\ldots
\varphi_{31}^+(t)
\varphi_{31}^-(t)
\varphi_{32}^+(t)
\varphi_1^+
\varphi_1^-
\ldots
\varphi_{32+n-1}^+
\varphi_{32+n-1}^-
\varphi_{32+n}^+\rangle
\nonumber\\
={\mbox {Pf}}
\left(
\begin{array}{cccccc}
 0         & \langle\varphi_1^+\varphi_1^-\rangle   & 
 \langle\varphi_1^+\varphi_2^+\rangle
  & \cdots & \langle\varphi_1^+(t)\varphi_{32+n}^+\rangle \\
 -\langle\varphi_1^+\varphi_1^-\rangle & 0          & 
 \langle\varphi_1^-\varphi_2^+\rangle
  & \cdots & \langle\varphi_1^-(t)\varphi_{32+n}^+\rangle \\
\vdots     & \vdots     & \vdots
  & \cdots & \vdots      \\
-\langle\varphi_1^+(t)\varphi_{32+n}^+\rangle& 
-\langle\varphi_1^-(t)\varphi_{32+n}^+\rangle
&-\langle\varphi_2^+(t)\varphi_{32+n}^+\rangle
  & \cdots & 0
\end{array}
\right) .
\nonumber
\end{eqnarray}
At last we
performed the integration over the time up to $t=1600$
with $\varepsilon =0.001$
and took into account in the sum over
$n$ up to $100$ neighbours.
The obtained dependences 
$S_{xx}(\kappa ,\omega )$ vs. $\omega$
for different
$\kappa$
at low temperature
and
$S_{xx}(0,\omega )$ vs. $\omega$
for few temperatures
are depicted in Fig. 1.
As it can be seen from Fig. 1 
$S_{xx}(\kappa,\omega)$
exhibits two peaks.
At $\beta=5$ and $\kappa=0$ one finds a high peak at 
$\omega_1=0.00$
and a low and broad one at 
$\omega_2=0.76$.
As $\kappa$ increases the 
height of the
first peak decreases and it shifts towards high
frequencies
($\omega_1 \approx 0.11$ at $\kappa =\frac{\pi}{4}$,
$\omega_1 \approx 0.25$ at $\kappa =\frac{\pi}{2}$,
$\omega_1 \approx 0.34$ at $\kappa =\frac{3\pi}{4}$,
$\omega_1 \approx 0.38$ at $\kappa =\pi$),
whereas the 
width of the 
second peak 
decreases, 
its height increases and it moves towards high frequencies
($\omega_2 \approx 0.80$ at $\kappa =\frac{\pi}{4}$,
$\omega_2 \approx 0.88$ at $\kappa =\frac{\pi}{2}$,
$\omega_2 \approx 0.96$ at $\kappa =\frac{3\pi}{4}$,
$\omega_2 \approx 1.01$ at $\kappa =\pi$).
As the temperature increases
$S_{xx}(0,\omega)$
qualitatively remains the same:
the heights of both peaks 
decrease and 
the high-frequency peak
shifts slightly towards high
frequencies
($\omega_2 \approx 0.76$ at $\beta=5$,
$\omega_2 \approx 0.77$ at $\beta=4$,
$\omega_2 \approx 0.81$ at $\beta=3$,
$\omega_2 \approx 0.86$ at $\beta=2$,
$\omega_2 \approx 0.91$ at $\beta=1$,
$\omega_2 \approx 0.95$ at $\beta=0.1$,
$\omega_2 \approx 0.96$ at $\beta=0.001$).
The discussed case of small transverse field
may be of interest for understanding the
dielectric measurement and neutron scattering data
for quasi-one-dimensional hydrogen-bonded ferroelectrics. 
However,
detailed comparison 
with experimental results
demands 
an introducing of weak interchain interactions (see [24,26])
that requires a separate study.

Next example regards to thermodynamics of
the isotropic $XY$ chain
($J_j^{xx}=J_j^{yy}=J_j$,
$J_j^{xy}=J_j^{yx}=0$)
with the random intersite couplings 
given, for example, by 
the Lorentzian probability distribution
density
$p(\ldots ,J_j,\ldots)=\prod_{j=1}^N
\frac{1}{\pi}\frac{\Gamma}{(J_j-J_0)^2+\Gamma^2}$
and the transverse fields that depend linearly on surrounding couplings
$\Omega_j-\Omega_0
=\frac{a}{2}(J_{j-1}+J_j-2J_0)$,
$\mid a\mid \ge 1$.
In Ref. [20] 
the random-averaged density of states 
$\overline{\rho (E)}$
was calculated 
that yielded thermodynamics of the model.
In particular, it was shown that 
the introduced randomness may cause the appearance
of non-zero averaged transverse magnetization
$\overline{m_z}=
-\frac{1}{2}
\int_{-\infty}^{\infty}{\mbox{d}}E
\overline{\rho (E)}\tanh{\frac{\beta E}{2}}$
at zero averaged transverse field
$\Omega_0=0$ 
($\overline{\rho (E)}$
is not symmetric with respect to the change
$E-\Omega_0\rightarrow -(E-\Omega_0)$
and at $T=0\;\;\;$
$-2\overline{m_z}=
\int_{-\infty}^{0}{\mbox{d}}E
\overline{\rho (E)}-
\int_{0}^{\infty}{\mbox{d}}E
\overline{\rho (E)}
\ne 0$).
We considered the described model for $N=15000$ with
$J_0=1$ for $a=\pm 1.01$ with 
the Lorentzian and Gaussian random couplings
(the probability distribution for the latter case reads
$p(\ldots ,J_j,\ldots)=\prod_{j=1}^N
\frac{1}{\sqrt{2\pi}\sigma}
\exp{
\left[
-\frac{(J_j-J_0)^2}{2\sigma^2}
\right]}
$)
and presented the results of computations for one random
realization in Fig. 2.
The displayed results are in complete agreement with the 
ones derived
analytically  
for the Lorentzian disorder.
However, the developed numerical procedure allows to study an arbitrary 
disorder.
The performed calculations have indicated the appearance of 
$\overline{m_z}\ne 0$
for $\Omega_0=0$ at low temperature for the correlated disorder.
This can be clearly demonstrated by the results of calculation of a number 
of negative and positive eigenvalues $\Lambda_k$ (2) 
${\cal N}_-$ and ${\cal N}_+$,
since at $T=0$ 
$-2\overline{m_z}=
\frac{{\cal N}_--{\cal N}_+}{N}$.
Putting $a=1.01$
for a certain Gaussian random realization with
$\sigma=0.25$($1$)
that yields
$\frac{1}{N}\sum_{j=1}^NJ_j=0.999757$
($0.999027$)
we found
${\cal N}_-=7192$($6024$) 
and
${\cal N}_+=7808$($8976$).
Another Gaussian random realization with
$\sigma=0.25$($1$)
that yields
$\frac{1}{N}\sum_{j=1}^NJ_j=1.000118$
($1.000473$) 
gave
${\cal N}_-=7187$($6073$) 
and
${\cal N}_+=7813$($8927$).
These results definitely point out the appearance of
"spontaneous magnetization" due to disorder.

We end up with the calculation of the $zz$ dynamic structure factor
$S_{zz} (\kappa ,\omega )
\equiv
\sum_{n=1}^N{\mbox {e}}^{{\mbox{i}}\kappa n}
\int_{-\infty}^{\infty}
$
$
{\mbox{d}}t
{\mbox {e}}^{-\varepsilon \mid t\mid}
{\mbox {e}}^{{\mbox{i}}\omega t}
\left(
\langle s_j^{z}(t)s_{j+n}^{z}\rangle
-\langle s_j^{z}\rangle\langle s_{j+n}^{z}\rangle
\right)
$
for the Ising chain
($J_j^{xx}=J=1$,
$J_j^{xy}=J_j^{yx}=J_j^{yy}=0$)
in the random transverse field defined by 
the probability distribution density
$p(\ldots ,\Omega_j,\ldots)=\prod_{j=1}^N
\left[
x\delta (\Omega_j)+(1-x)\delta (\Omega_j-0.5) 
\right] 
$,
$0\le x\le 1$.
We computed the correlation functions
\linebreak
$4\langle s^z_{100}(t)s^z_{100+n}\rangle
=\langle\varphi_{100}^+\varphi_{100}^-\rangle
\langle\varphi_{100+n}^+\varphi_{100+n}^-\rangle-
\langle\varphi_{100}^+(t)\varphi_{100+n}^+\rangle
\langle\varphi_{100}^-(t)\varphi_{100+n}^-\rangle+
\langle\varphi_{100}^+(t)\varphi_{100+n}^-\rangle
$
$
\langle\varphi_{100}^-(t)\varphi_{100+n}^+\rangle$
for $250$ random chains of $200$ spins,
performed the integration over time $t$
with $\varepsilon=0.005$
and the summation over neighbours $n$.
The obtained random-averaged $zz$ dynamic structure factor that
for $\kappa=0$ is
presented in Fig. 3. 
The depicted plots demonstrate how the frequency-dependent 
$zz$ structure factor rebuilds from 
the Ising type behaviour to the transverse 
Ising type behaviour as the concentration of sites with transverse field 
increases from $0$ ($x=1$) to $1$ ($x=0$). The obtained 
dynamic structure factor exhibits a lot of structure that is induced by 
the disorder arrangement of two values of transverse field $0$ and $0.5$.
It appears that each well-defined peak for small concentrations of $1-x$ is 
connected with $S_{zz} (0,\omega )$ for a certain chain determined by local 
environment of spin at $j=100$
(for example,
$\ldots 0{\bf {0}}0\ldots,\;$
$\ldots 0\Omega {\bf {0}}0\ldots,\;$
$\ldots 0{\bf {\Omega}}0\ldots,\;$
$\ldots 0\Omega \Omega {\bf {0}}0\ldots,\;$
$\ldots 0 \Omega {\bf {\Omega}}0\ldots,\;$
$\ldots 0 \Omega {\bf {0}}\Omega \ldots\;$
etc., the transverse field $\Omega_{100}$ is written in bold font,
not written $\Omega_j$s do not influence
$S_{zz} (0,\omega )$).
With decreasing of $x$ a number of possible local structures
in the vicinity of $j=100$
(and thus a number of peaks) increases and the peaks appear almost at all 
frequencies. However, the difference in their heights is conditioned by the 
probability of their appearance that is large.
As a result one gets fine structure that transforms into 
the smooth curve only 
in the limiting case $x=0$.
The described random model has a simple interpretation in connection with  
partially deuterated quasi-one-dimensional hydrogen-bonded ferroelectrics.
However, a study of relevant there $xx$ spin dynamics 
is more cumbersome and will be reported 
separately.

In summary, we have presented the numerical approach 
suitable for
calculation of time-dependent correlation functions for 
non-random and random 
spin-$\frac{1}{2}$
$XY$ chains. 
We have illustrated the 
numerical procedure deriving some new results.
It is relevant to mention here the papers [36,37]
devoted to numerical calculations of 
the $xx$ time-dependent spin correlation 
functions for the isotropic $XY$ model.
The authors used explicit expressions for 
the elementary contractions
(in contrast to our formulae that also fit the random models)
and computed determinants of corresponding antisymmetric matrices that 
yielded only the square of correlation functions
(this causes some difficulties in 
further calculation of the dynamic structure factor 
or susceptibility as well as in the
study of random models).
We hope that the numerical analysis of the properties of spin-$\frac{1}{2}$
$XY$ chains will be useful for understanding the results of the 
corresponding measurements on quasi-one-dimensional hydrogen-bonded 
ferroelectrics like
CsH$_2$PO$_4$,
Cs(H$_{1-{\mbox{x}}}$D$_{\mbox{x}}$)$_2$PO$_4$,
PbHPO$_4$,
PbH$_{1-{\mbox{x}}}$D$_{\mbox{x}}$PO$_4$
(neutron scattering, dielectric measurement)
and $J$-aggregates (absorption and emission spectra).
However, a comparison of theoretical predictions and experimental data 
requires further studies.

\clearpage

\clearpage

\clearpage

\clearpage

\noindent
{\bf List of figure captions}\\

\vspace{2.25cm}

\noindent
FIG. 1.
The dynamic structure factor for the
Ising chain in transverse field.
(a)
$S_{xx}(\kappa ,\omega)$ vs. $\omega$
at $\beta=5$ for 
$\kappa =0,
\frac{\pi}{4},
\frac{\pi}{2},
\frac{3\pi}{4},
\pi$
(curves $1,\ldots ,5$, respectively)
and
(b)
$S_{xx}(0,\omega)$ vs. $\omega$
for $\beta =5,
4,
3,
2,
1,
0.1,
0.001$
(curves $1,\ldots ,7$, respectively).

\vspace{2.25cm}

\noindent
FIG. 2.
The density of states and transverse magnetization for the
isotropic $XY$ chain with random couplings and transverse fields that 
depend linearly on surrounding couplings.
$\rho (E)$ vs. $E-\Omega_0$ (a,c)
and
$-\overline{m_z}$ vs. $\Omega_0$ at $\beta =100$ (b,d)
for the Lorentzian (short dashed lines)
and Gaussian (solid lines)
disorder
with 
$\Gamma =\sigma =0.25$ (a,b)
and
$\Gamma =\sigma =1$ (c,d);
the results for non-random case
$\Gamma =\sigma =0$ 
are depicted by long dashed lines.

\vspace{2.25cm}

\noindent
FIG. 3.
The frequency dependence of
the averaged dynamic structure factor $\overline{S_{zz}(0,\omega)}$ 
for the Ising chain in random transverse field
at $\beta=5$.

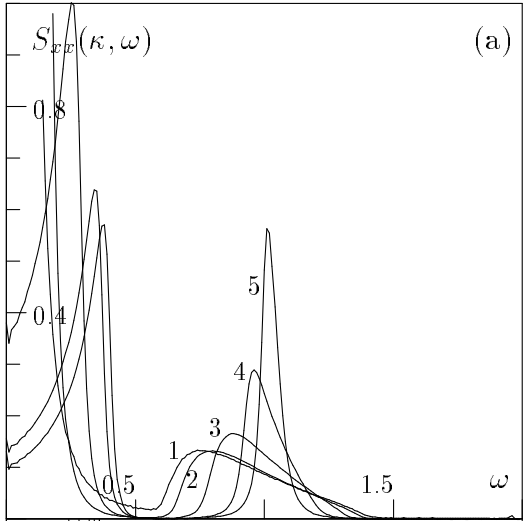
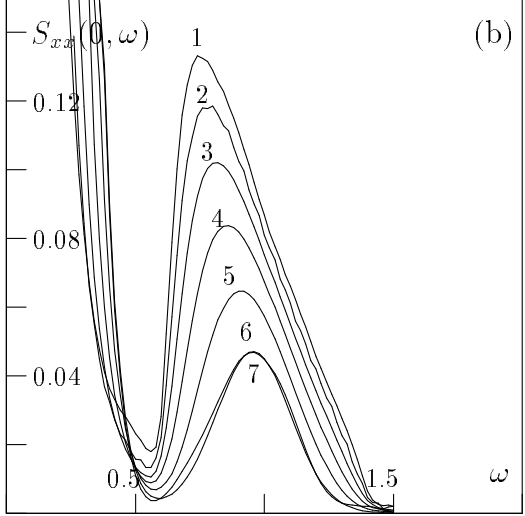
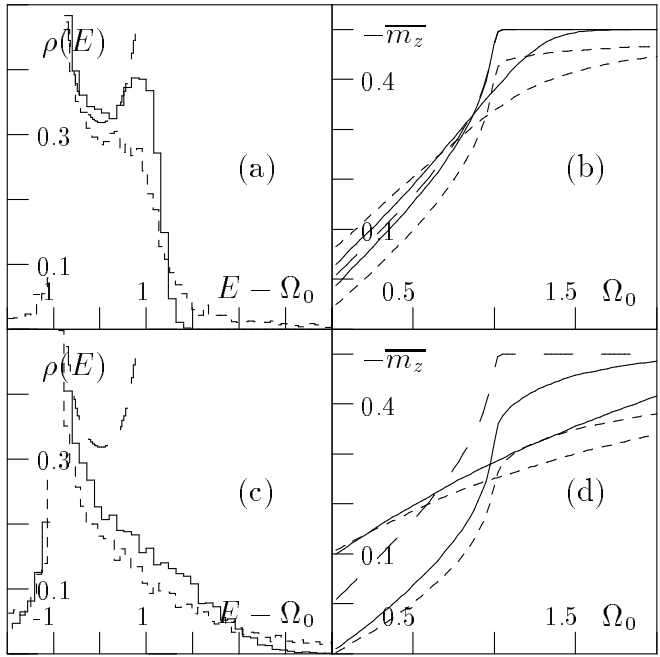
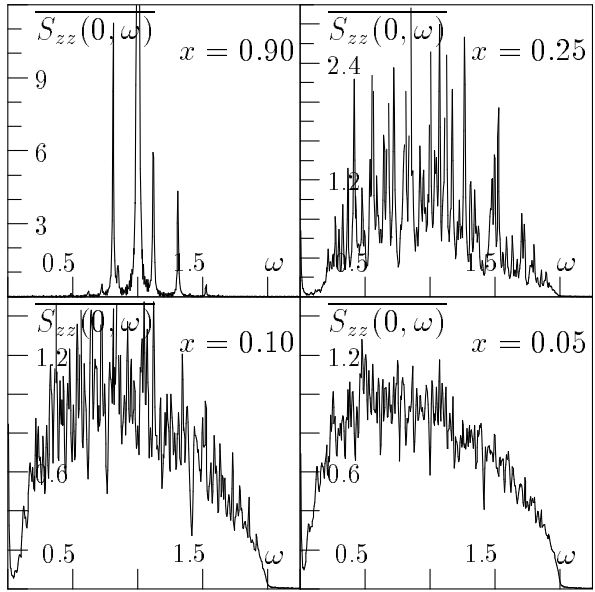
\end{document}